\documentclass[acmsmall,screen]{acmart}

\setcopyright{acmlicensed}
\copyrightyear{2025}
\acmYear{2025}
\acmDOI{XXXXXXX.XXXXXXX}

\PassOptionsToPackage{dvipsnames,table}{xcolor}
\usepackage{xcolor}
\usepackage{booktabs} 
\usepackage{tabularx}
\usepackage{colortbl}
\usepackage{arydshln}
\usepackage{adjustbox}
\usepackage[linesnumbered,ruled,vlined]{algorithm2e}
\usepackage{amsmath}
 
\usepackage{amssymb}
\usepackage{bm}
\usepackage{enumitem}
\usepackage{soul}
\usepackage{threeparttable}  
\usepackage{multirow}
\usepackage{makecell}
\usepackage{tabularx}
\usepackage[T1]{fontenc}
\usepackage{array}
\usepackage{listings}
\usepackage{xspace} 
\usepackage{subcaption}
\usepackage{graphicx}
\usepackage{hyperref}
\usepackage{textcomp}
\usepackage{microtype} 
\usepackage{calc}
\usepackage{pifont}
\usepackage{tcolorbox}
\usepackage{lscape}
\usepackage{rotating}
\usepackage{tikz}
\usepackage{mdframed}
\usetikzlibrary{positioning}
\usepackage{balance}

\newcolumntype{Y}{>{\centering\arraybackslash}X}
\newcolumntype{Z}{>{\raggedright\arraybackslash}X}

\SetCommentSty{mycommfont}
\SetKwInput{KwInput}{Input}                
\SetKwInput{KwOutput}{Output}              

\newlength{\ColorBoxDepthReference}
\newlength{\ColorBoxHeightReference}
\newlength{\Width}%
\newcommand{\MyColorBox}[2][red]%
{%
	\settowidth{\Width}{#2}%
	\colorbox{#1}%
	{%
		\raisebox{-\ColorBoxDepthReference}%
		{%
			\parbox[b][\ColorBoxHeightReference+\ColorBoxDepthReference][c]{\Width}{\centering#2}%
		}%
	}%
}
\definecolor{codegreen}{rgb}{0,0.6,0}
\definecolor{codegray}{rgb}{0.5,0.5,0.5}
\definecolor{codepurple}{rgb}{0.58,0,0.82}

\newcommand{\parabf}[1]{\noindent\textbf{#1}}

\newcommand{\oppotunity}[2]{
 	\vspace{1mm}
	\begin{mdframed}[linecolor=gray,roundcorner=12pt,backgroundcolor=gray!15,linewidth=3pt,innerleftmargin=2pt, leftmargin=0cm,rightmargin=0cm,topline=false,bottomline=false,rightline = false]
		\textbf{\textit{Oppotunity-#1:}} 
        
            #2
	\end{mdframed}
	\vspace{3mm}
}

\newcommand{\codedigitaltwin}{\textsc{Code Digital Twin}\xspace}
\newcommand{\sysdigitaltwin}{\textsc{System Digital Twin}\xspace}

\begin{document}

\title{\codedigitaltwin: A Knowledge Infrastructure for AI-Assisted Complex Software Development}

\author{Xin Peng}
\email{pengxin@fudan.edu.cn}
\affiliation{\institution{College of Computer Science and Artificial Intelligence, Fudan University}\country{China}}

\author{Chong Wang}
\email{wangchong20@fudan.edu.cn}
\affiliation{\institution{College of Computer Science and Artificial Intelligence, Fudan University}\country{China}}


\begin{abstract}
Recent advances in AI coding tools powered by large language models (LLMs) have shown strong capabilities in software engineering tasks, raising expectations of major productivity gains.
Tools such as Cursor and Claude Code have popularized ``vibe coding'' (where developers steer development through high-level intent), commonly relying on \textit{context engineering} and Retrieval-Augmented Generation (RAG) to ground generation in a codebase.
However, these paradigms struggle in ultra-complex enterprise systems, where software evolves incrementally under pervasive design constraints and depends on \textbf{\textit{tacit knowledge}} such as responsibilities, intent, and decision rationales distributed across code, configurations, discussions, and version history.
In this environment, context engineering faces a fundamental barrier: the required context is scattered across artifacts and entangled across time, beyond the capacity of LLMs to reliably capture, prioritize, and fuse evidence into correct and trustworthy decisions, even as context windows grow.
To bridge this gap, we propose the \textbf{\codedigitaltwin}, a persistent and evolving \textbf{knowledge infrastructure} built on the codebase. It separates \textit{long-term knowledge engineering} from \textit{task-time context engineering} and serves as a backend ``context engine'' for AI coding assistants.
The \codedigitaltwin models both the physical and conceptual layers of software and co-evolves with the system. By integrating hybrid knowledge representations, multi-stage extraction pipelines, incremental updates, AI-empowered applications, and human-in-the-loop feedback, it transforms fragmented knowledge into explicit and actionable representations, providing a roadmap toward sustainable and resilient development and evolution of ultra-complex systems.
\end{abstract}



\received{2025}
\received[revised]{2025}
\received[accepted]{2025}

\settopmatter{printfolios=true} 

\maketitle

\section{Introduction}
Recent advances in large language models (LLMs) have enabled a new generation of AI coding assistants, including Cursor~\cite{cursor}, GitHub Copilot~\cite{copilot}, and Claude Code~\cite{claudecode}, which achieve strong results on public software engineering (SE) benchmarks and leaderboards for tasks such as code generation and issue resolution. These assistants, along with the emerging ``vibe coding'' paradigm, help with routine development activities by understanding and generating code, and they can improve productivity for well-scoped tasks when grounded in relevant project context. These successes have also given rise to ambitious claims such as ``generative AI ending programming'' and ``AI revolutionizing software paradigms.'' Such hype can mislead enterprise leaders into expecting ``10$\times$ efficiency gains,'' which in turn creates unrealistic demands on development teams. As a result, research and industry efforts often prioritize the visible ``glamour effect'' of model and agent capabilities, while paying limited attention to the intrinsic challenges of long-lived, complex software evolution.

Enterprise software increasingly serves as critical foundation, and its scale and complexity often grow into ultra-complex systems. Over time, countless changes accumulate into layered design trade-offs and conventions that span from requirements through architecture to implementation, making each system's evolutionary trajectory highly idiosyncratic. For instance, the Linux kernel has expanded by more than 10$\times$ over two decades (now exceeding 30M lines of code), reflecting successive optimizations, compromises, and design rationales. Likewise, microservice architectures can evolve into sprawling ecosystems whose interdependencies are difficult to manage. Given this nature, enterprise software engineering is a continuous process of development and evolution; even ``adding a new feature'' rarely starts from scratch. Changes must fit within accumulated spatial and temporal constraints, including dependencies, non-functional requirements, and long-standing design consistency. Engineers must interpret requirements, locate relevant concerns, understand existing code, devise solutions, and assess downstream impacts to preserve system integrity. Doing this reliably depends on \textit{tacit knowledge} behind software (e.g., high level concepts, functionalities, design rationales, and historical decisions), which is seldom documented and often decays over time. These gaps limit both developers and AI coding agents, forcing repeated reconstruction of context when solving development problems.

\begin{figure}[t]
\centering
\includegraphics[width=\linewidth]{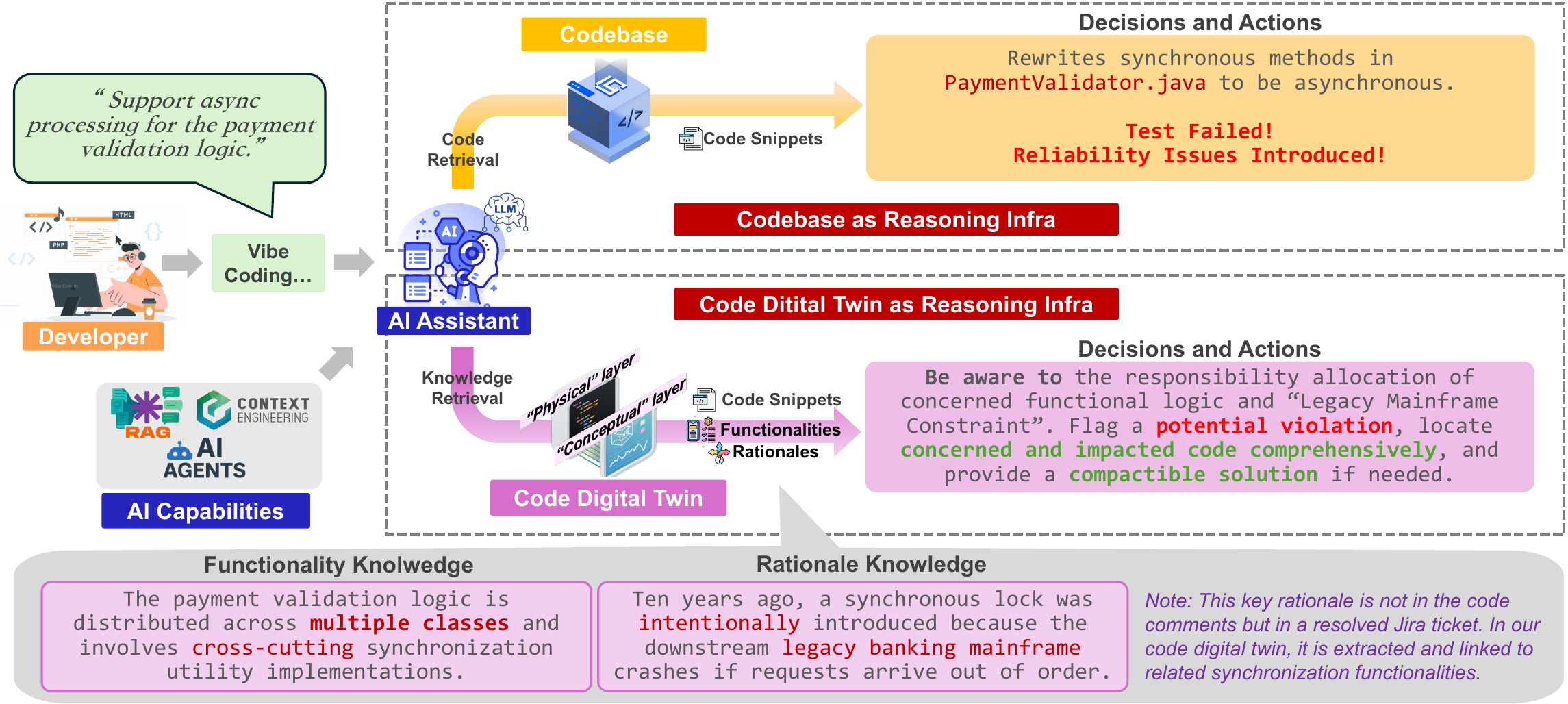}
\caption{Illustration of the vibe-coding trap under code-centric context engineering, and how \codedigitaltwin provides a knowledge infrastructure for smarter AI-assisted development in long-lived systems.}\label{fig:motivation}
\label{fig:motivation-vibe-coding}
\end{figure}

\textbf{The Context Engineering Bottleneck.} Modern AI coding tools rely heavily on \textit{context engineering}, the practice of selecting and organizing relevant project information (code and documentation snippets) to provide to the model. This paradigm enables ``vibe coding,'' allowing developers to focus on intended behavior rather than low-level syntax. However, it implicitly assumes that the retrieved code context captures all necessary truth. In complex, long-lived systems, code is the endpoint of evolving requirements, constraints, and trade-offs, and the underlying \textit{tacit knowledge} is often distributed across modules, documents, commits, issues, and discussions. Consequently, as illustrated in Figure~\ref{fig:motivation}, \textit{code-centric} retrieval and context engineering can (partially) recover \textit{what} the code does today, but they often miss \textit{how} it evolved and \textit{why} it took its current form. Unlike general question answering, where an answer may reside in a small number of retrieved documents, software knowledge is scattered spatially across coupled modules and entangled temporally across change histories, making it difficult to assemble accurate and complete context for nontrivial tasks. This limitation becomes a critical bottleneck for AI-assisted enterprise development and maintenance, which we term \textit{uncontrollable knowledge entropy}.

In particular, we identify 11 challenges for complex software development in the AI era, spanning both software-system and AI-assistant perspectives. From the software-system perspective, challenges stem from intrinsic complexity, the lack of explicit conceptual representations, long-term codebase evolution, undocumented knowledge loss, and socio-technical dependencies. These issues are rooted in architecture, organizational practices, and historical trajectories, and they define the constraints that any AI-assisted approach must respect. From the AI-assistant perspective, challenges focus on making AI contribute effectively, including precise task formalization, context-aware reasoning, accounting for non-functional and operational constraints, producing trustworthy outputs under human oversight, assistant/agent design, and realistic evaluation. Alongside these challenges, we highlight opportunities such as capturing implicit knowledge, building structured and dynamic representations, supporting impact analysis, enhancing collaboration, and leveraging AI as augmented intelligence. Together, these perspectives highlight a central tension: gaps in systems' \textit{tacit knowledge} limit what AI assistance can do effectively and safely, yet AI can also help elicit and operationalize that tacit knowledge to amplify human expertise while managing risk in long-lived development and maintenance. This motivates an infrastructure layer built on top of the codebase to capture, maintain, and serve tacit knowledge as actionable context for reliable context engineering and AI-assisted coding.

We propose \textit{\textbf{\codedigitaltwin}}, a living knowledge infrastructure that captures, structures, and continuously updates tacit knowledge across multiple levels, bridging evolving software systems and AI assistants. It separates \textit{long-term knowledge engineering} from \textit{task-time context engineering} through a systematic methodology, enabling the sustained capture, accumulation, and reuse of tacit knowledge. Drawing on the digital-twin paradigm in manufacturing (a coupled physical system and its digital representation), \codedigitaltwin couples a ``physical'' layer of software artifacts (e.g., functions, files, and modules) with a ``conceptual'' layer that encodes the system's intent and evolution, including domain concepts and features, their relationships and constraints, responsibility allocation across components, architectural dependencies, contextual decisions, and the design rationales and trade-offs behind them. The two layers are connected through bidirectional, traceable links to concrete artifacts and their histories (e.g., versions, commits, issues, and discussions), so that the twin remains consistent and updatable as the code changes. This mapping enables co-evolution and mitigates tacit knowledge decay: the \textit{artifact-oriented backbone} grounds high-level concepts and functionalities in concrete artifacts, while \textit{rationale-centric explanations} preserve the decisions, constraints, and trade-offs that shaped system design and evolution. As an infrastructure layer, complemented by an upper \textit{AI capability layer} (e.g., tailored RAG, context engineering, memory management, and agentic algorithms), \codedigitaltwin enables developers and AI assistants to assemble the necessary context, gain comprehensive insights, explore alternative solutions, and assess impacts and risks, thereby supporting complex engineering decisions\footnote{For online running systems (e.g., cloud-native systems), it naturally extends to \textit{\textbf{\sysdigitaltwin}} by incorporating deployment states, configurations, and operational data, enabling similar support.}. Figure~\ref{fig:motivation} illustrates how \codedigitaltwin leverages tacit knowledge to guide an AI assistant during refactoring in a long-lived system.

Specifically, \codedigitaltwin is realized through a hybrid knowledge stack that combines (i) structured representations (e.g., knowledge graphs, frames, and cards) to encode concepts, features, responsibilities, constraints, and rationales, and (ii) unstructured representations to preserve supporting textual evidence and nuance that complements the structured backbone. Its construction pipeline extracts artifact-oriented backbone knowledge from code and documentation, captures rationale-centric knowledge from unstructured sources such as commits, issues, and mailing lists, and establishes bidirectional links between knowledge and artifacts to enable traceability and context-aware analysis. Continuous co-evolution keeps the twin aligned with ongoing changes in artifacts, functionality, and rationale. Building on this foundation, the AI capability layer supports knowledge exploration and context-aware assistance, while human-in-the-loop mechanisms validate and steer updates. Our roadmap highlights research needs in scalable hybrid representations, higher-fidelity extraction, near-real-time synchronization with codebases, tighter human-AI collaboration, and adaptive interfaces so that the twin remains accurate, comprehensive, and actionable for long-term software engineering. 

In summary, our contributions are detailed as follows:
\begin{itemize}
\item \textit{Diagnosing the context engineering bottleneck for enterprise software:} We analyze why code-centric retrieval and ad hoc \textit{context engineering} are insufficient for long-lived, complex systems, and we formulate this limitation as \textit{unmanaged context entropy}. We further summarize challenges and opportunities for making AI assistance effective and trustworthy under evolving architectural, historical, and socio-technical constraints.
\item \textit{Introducing \codedigitaltwin as a living knowledge infrastructure:} We propose \textbf{\codedigitaltwin} to capture and maintain enterprise \textit{tacit knowledge} alongside evolving software artifacts. \codedigitaltwin couples an artifact layer (code and related artifacts) with a conceptual layer (concepts, responsibilities, constraints, and rationales), connected via bidirectional, traceable links to support version-aware, updatable knowledge.
\item \textit{Outlining a construction methodology and research roadmap:} We present a hybrid knowledge stack and a construction pipeline that extracts code and artifact map, functionality-oriented skeleton knowledge, and rationale-centric explanatory knowledge from heterogeneous sources, links them for traceability, and supports continuous co-evolution with code changes. We also outline a roadmap for scalable representations, higher-fidelity extraction, incremental synchronization, human-in-the-loop validation, and interfaces that enable context-aware assistance.
\end{itemize}

\section{Motivation}
Recent advances in AI coding assistants have enabled tools that perform well on public software engineering benchmarks such as SWE-bench~\cite{jimenez2023swe}. However, benchmark success does not directly translate to reliable support for large, mission-critical, long-lived systems. Complex software systems, such as the Linux kernel with hundreds of thousands of contributors and over a million annual commits~\cite{linux}, require more than local code edits: developers must preserve architectural intent, interpret historical decisions, and reason across distributed dependencies. This section clarifies what AI coding assistants can improve in day-to-day work and what they do not resolve in complex system evolution.

\subsection{What AI Coding Assistants Change}
AI coding assistants primarily improve localized and routine development activities. They help draft boilerplate code, generate experimental prototypes, and translate between programming languages, reducing manual effort for well-scoped edits. They also support code understanding by summarizing code snippets and producing readable explanations, API summaries, and usage examples~\cite{sun2024source,dvivedi2024comparative}.

Beyond reading and documentation, these assistants can support testing and quality assurance by proposing unit tests, repairing simple failing tests, and expanding coverage for bounded code portions~\cite{schafer2023empirical,dakhel2024effective}. They can also help with issue resolution~\cite{jimenez2023swe}, pull request review, and other repetitive coordination tasks by offering drafts that developers can validate and refine.

\subsection{What AI Coding Assistants Do Not Change}
Despite these advances, the core difficulties of software engineering remain. Large, long-lived systems accumulate tightly coupled dependencies, external operating constraints, and decades of decentralized evolution. As a result, both maintenance (e.g., diagnosing failures and planning safe changes) and development (e.g., implementing features iteratively) require system-wide reasoning and the integration of explicit artifacts with \textit{tacit knowledge}.
These challenges reflect Brooks' \textit{\textbf{essential complexity of software}}: the inherent difficulty of conceptualizing, designing, and coordinating large, evolving systems that cannot be removed by automation alone~\cite{brooks1987no}.

Tacit knowledge, including architectural rationales, design trade-offs, and historical context, often lives in developer experience or informal artifacts rather than in code. Current assistants cannot reliably retrieve or reconstruct this knowledge on demand. Empirical studies show persistent weaknesses in system-wide reasoning~\cite{lancer}, long-horizon analysis~\cite{deng2025swebenchproaiagents}, and architectural-level decision making. As a result, capability gains alone do not eliminate the need for structured, version-aware context that preserves why the system is the way it is.

\subsection{Our Vision}
AI coding assistants can be useful collaborators for complex systems when their suggestions are grounded in accurate, system-specific context. The practical goal is not to replace human judgment, but to reduce the cost of routine work and to make system constraints and rationales easier to surface during change planning.

This paper argues that achieving this reliability requires a persistent, curated knowledge infrastructure layer that captures and maintains \textit{tacit knowledge} alongside evolving artifacts, rather than relying on ad hoc retrieval at query time. Later sections introduce \codedigitaltwin as a step in this direction and outline how it can support context-aware assistance under human oversight.

\textbf{Illustrative Scenario: The `Vibe Coding' Trap in Legacy Systems.}
Consider a developer using an AI assistant to ``refactor the payment validation logic to support async processing'' as illustrated in Figure~\ref{fig:motivation}.
\begin{itemize}
    \item \textbf{Standard AI coding assistant:} The tool scans the current \texttt{PaymentValidator.java}, sees synchronous methods, and rewrites them to be asynchronous. It produces syntactically correct code aligned with the requested behavior.
    \item \textbf{The Hidden Failure:} Ten years ago, a synchronous lock was intentionally introduced because the downstream legacy banking mainframe crashes if requests arrive out of order. This rationale is not in the code comments but in a resolved Jira ticket.
    \item \textbf{\codedigitaltwin approach:} \codedigitaltwin links \texttt{PaymentValidator} not only to its code, but also to the historical rationale entity ``Legacy Mainframe Constraint.'' When the assistant proposes async processing, \codedigitaltwin can surface this constraint and flag a potential violation.
\end{itemize}

\section{Complex Software Development in the AI Era: Challenges and Opportunities}
In this section, we discuss key challenges and opportunities of complex software development in the AI era, particularly with the advent of AI coding assistants.

\subsection{Overview: Bridging Software-System and AI-Assistant Perspectives}
Complex software development in the AI era can be understood through two complementary lenses: the \textit{software-system perspective} and the \textit{AI-assistant perspective}. Table~\ref{tab:opportunities} summarizes the key challenges and corresponding actionable opportunities identified from both perspectives.

From the \textit{software-system perspective}, challenges arise from the intrinsic complexity of building and evolving software systems, the lack of explicit conceptual representations, the historical accumulation of changes, the loss of undocumented knowledge, and socio-technical dependencies. These challenges are deeply rooted in system architecture, organizational practices, and long-term evolution, and they define the constraints and requirements that any automated or AI-assisted approach must respect across both development and maintenance.
The \textit{AI-assistant perspective}, by contrast, focuses on how AI tools can support developers in creating, extending, and sustaining software, highlighting both limitations and opportunities. AI coding assistants can automate repetitive coding tasks, provide drafts for debugging or refactoring, and support compliance or performance checks. Their usefulness depends on precise task specification, access to system-specific context, respect for non-functional and operational constraints, and outputs that developers can validate. In practice, these tools must fit into developer workflows, incorporate domain knowledge, and account for the historical and architectural uniqueness of complex codebases.

The relationship between these perspectives is bidirectional. Software challenges in both development and maintenance motivate AI tool design, shaping which capabilities assistants should provide. For example, context-aware reasoning is needed to implement new features without breaking dependencies, while assistant design can reduce repetitive coordination overhead. Conversely, current limitations of AI assistants reinforce long-standing requirements for human oversight, careful task specification, and explicit representations of \textit{tacit knowledge}. Together, these perspectives provide a unified view: the structural, historical, and organizational complexity of software sets the conditions under which AI tools can be safely applied, while AI techniques can reduce routine effort and help surface relevant context during development and maintenance.

\begin{table}
\centering
\scriptsize
\renewcommand{\arraystretch}{1.2}
\caption{Challenges and Actionable Opportunities from Software-System and AI-Assistant Perspectives}
\label{tab:opportunities}
\begin{tabularx}{\textwidth}{c>{\raggedright\arraybackslash}p{2cm}>{\raggedright\arraybackslash}X>{\raggedright\arraybackslash}X}
\toprule
\textbf{Perspective}      & \textbf{Challenges}                              & \textbf{Actionable Opportunities}                                                                                                                                                                       & \textbf{Examples / Notes}                                                                                                                              \\ 
\midrule
\multirow[c]{6}{*}{\textbf{Software-System}} & Intrinsic System Complexity                & Model dependencies across code, build, and runtime boundaries; link them to configuration and environment constraints; support change-impact analysis that estimates affected components and regression surfaces. & A banking system could automatically highlight all modules affected by a business rule update across UI, database, and service layers.                 \\ 
\cline{2-4}
                          & Physical vs. Conceptual Representation             & Extract architectural roles, workflows, and constraints from code and related artifacts; link conceptual elements to concrete files, APIs, and tests; make implicit invariants checkable during development and maintenance.                                     & In a healthcare system, map scattered privacy enforcement logic across modules into a central knowledge structure to support compliance verification.  \\ 
\cline{2-4}
                          & Evolutionary Path Uniqueness                    & Build evolution-aware traces that capture what changed and why; identify hotspots and fragile zones using historical signals; generate history-sensitive change guidance that respects compatibility constraints.                                         & ERP systems could identify which legacy modules should not be refactored due to backward compatibility or regulatory reasons.                          \\ 
\cline{2-4}
                          & Undocumented Knowledge Loss                           & Extract rationales and assumptions from development traces; provide lightweight capture points at decision-heavy moments; curate knowledge with provenance and freshness to keep context usable over time.                                                          & A temporary performance workaround documented in a code review can be explicitly linked to its rationale for future maintainers.                       \\ 
\cline{2-4}
                          & Socio-Technical and Organizational Dependencies & Link technical artifacts to ownership and cross-team interfaces; detect coordination needs early and reflect them in planning workflows; recommend sequencing and communication checkpoints for multi-party changes.                                                  & Deprecating a shared API triggers notifications for all dependent teams, along with suggested migration steps.                                         \\ 
\midrule
\multirow[c]{6}{*}{\textbf{AI-Assistant}}      & Task Specification                              & Represent tasks with explicit intent, inputs/outputs, constraints, and acceptance criteria; support interactive elicitation from informal requests; attach validation hooks before code is proposed or changed.                              & Translating ``optimize payment module'' into specific actions with transaction atomicity, input validation, and integration checks.                      \\ 
\cline{2-4}
                          & Context Gap and Context-Aware Reasoning                         & Move from text snippets to dependency-aware context resolution; implement systematic context selection and filtering with explicit signals; use version-aware traces to surface what changed and why.                                                                          & When updating a shared utility function, the system identifies all downstream modules and hidden contracts to prevent regressions.                     \\ 
\cline{2-4}
                          & Non-Functional and Operational Constraints      & Represent policies and budgets as explicit constraints; integrate pre-flight validation (tests, analyses, compliance checks); connect operational signals to constraint refinement to reduce regressions.                                  & Querying ``optimize database query'' will consider memory limits, access controls, and response time constraints.                                               \\ 
\cline{2-4}
                          & Trust and Human Oversight                       & Provide provenance and evidence for suggestions; communicate expected impact in actionable terms; integrate staged verification, rollout, and rollback into assistant-supported changes.                                             & An assistant proposes a refactoring with linked rationale, affected modules, and rollback instructions for human review.                                          \\ 
\cline{2-4}
                          & Intelligent Assistant Design                    & Combine artifact reasoning with tool integration for iterative verification; maintain version-aware project memory to reduce repeated context engineering; support cross-artifact reasoning for safe refactoring and migration decisions.                                                          & An API migration advisor that considers module dependencies, usage patterns, backward compatibility, and test coverage.                                \\ 
\cline{2-4}
                          & Realistic Evaluation and Benchmarking           & Build datasets that include historical evolution and multi-module dependencies; design tasks that require preserving constraints, not only producing patches; evaluate longitudinally to test robustness under version and requirement changes.          & Assess an assistant's ability to fix a decade-old banking system bug while preserving regulatory compliance and module interdependencies.                    \\
\bottomrule
\end{tabularx}
\end{table}

\subsection{Complex Software Development: Software-System Perspective}
Complex software development is a continuous process that integrates both building new capabilities and maintaining existing ones, ensuring long-term reliability, security, and adaptability. In the AI era, these activities face inherent challenges from system complexity, historical evolution, and organizational factors, while also opening new opportunities through AI coding assistants.

\textit{\textbf{Challenge I: Intrinsic System Complexity.}}  
Internally, software often exhibits high coupling, cross-cutting concerns, tangled dependencies, architectural drift, and poor separation of concerns (for example, UI logic intermingled with business logic or data access code embedded in service layers). A typical example is a legacy banking system where business rules are scattered across presentation and database layers, making it difficult to isolate logic for compliance updates. Externally, software must interact with third-party libraries, legacy APIs, operating systems, hardware platforms, networks, concurrency mechanisms, security policies, runtime environments, deployment infrastructures, and diverse user contexts. For instance, a mobile application may work on Android but fail on iOS due to subtle API differences. These internal and external factors introduce additional constraints and unpredictability. In practice, assistants have limited visibility into architecture and dependencies that are not explicit in the local code view (for example, runtime configurations, version constraints, and deployment-specific behavior), which increases the risk of missing side effects.

\oppotunity{I}{
\begin{itemize}[label=-,leftmargin=10pt]
    \item Maintain a continuously updated dependency model that covers code-level calls, build-time links, and runtime interactions across modules and external services.
    \item Connect the dependency model to configuration, deployment, and environment artifacts so that behavior differences across platforms and versions become explicit.
    \item Provide query and visualization support for exploring coupling, cross-cutting concerns, and architectural drift at multiple granularities.
    \item Offer change-impact analysis that predicts affected components, integration points, and likely regression surfaces before modifications are applied.
\end{itemize}
}

\textit{\textbf{Challenge II: Physical vs. Conceptual Representation.}}  
Source code serves as the physical artifact in the form of text, files, and modules, but the conceptual representation of a system, its architectural layers, module responsibilities, workflows, and non-codified constraints, remains largely implicit. For example, in a healthcare system, logic enforcing patient privacy rules may be scattered across multiple modules and undocumented, making it difficult to verify compliance or adapt to new regulations. Design rationales, domain models, data flow diagrams, business rules, and invariants are often incomplete or missing in older systems. As a result, tasks such as refactoring, debugging, or feature integration frequently require reconstructing these conceptual models. AI assistants struggle to infer hidden invariants, implicit business rules, or evolving requirements that are not explicitly reflected in code.

\oppotunity{II}{
\begin{itemize}[label=-,leftmargin=10pt]
    \item Extract conceptual elements (architectural roles, workflows, data contracts, business rules) from code, tests, and documentation, and represent them in a stable schema.
    \item Link conceptual elements to concrete artifacts (files, APIs, configurations, tests) so that traceability supports both navigation and verification.
    \item Make implicit constraints explicit (invariants, assumptions, policy rules), and expose them as checkable conditions during development and maintenance.
    \item Detect and report representation gaps (missing rationale, incomplete documentation, unowned constraints) to guide targeted knowledge capture.
\end{itemize}
}

\textit{\textbf{Challenge III: Evolutionary Path Uniqueness.}}  
Each software system carries the imprint of its history, including its origin, successive design trade-offs, accumulated patches, contributions from different developers, refactorings, and platform migrations. For instance, an enterprise resource planning (ERP) system that began as a small accounting tool may, over decades, accumulate inconsistent naming conventions, duplicated modules, temporary fixes that became permanent, and obsolete patterns retained for backward compatibility. This historical accumulation makes each system resistant to uniform solutions. Assistants can miss locally important conventions and compatibility constraints that determine which changes are safe.

\oppotunity{III}{
\begin{itemize}[label=-,leftmargin=10pt]
    \item Record an evolution graph that captures what changed, when it changed, and the associated rationale across commits, issues, and reviews.
    \item Identify historical hotspots and fragile zones by analyzing change frequency, defect density, and dependency centrality over time.
    \item Characterize legacy structures and technical-debt patterns as reusable categories, enabling consistent triage and refactoring planning.
    \item Generate change recommendations that are history-sensitive, highlighting compatibility constraints, preserved behavior, and prior failed attempts.
\end{itemize}
}

\textit{\textbf{Challenge IV: Undocumented Knowledge Loss.}}  
Over time, critical knowledge about a system can be lost not only when developers and architects leave the organization, but also because key decisions are never formally documented. This includes insights shared in unrecorded meetings, informal discussions among team members, and lessons learned through hands-on experience while fixing issues or resolving incidents. Such undocumented knowledge encompasses design rationales, hidden assumptions, operational workarounds, and contextual understanding of complex system behavior. Its loss leaves future maintainers without essential context, increasing the likelihood of errors or invariant violations when making changes. For example, a ``temporary'' performance shortcut or an unrecorded dependency discovered during a past bug fix may be essential for production stability but invisible to new maintainers. AI assistants cannot reliably recover this institutional memory or infer the unstated reasoning embedded in historical practices and informal artifacts.

\oppotunity{IV}{
\begin{itemize}[label=-,leftmargin=10pt]
    \item Extract knowledge from existing development traces (commits, issues, reviews, incident notes) and normalize it into structured, linkable records.
    \item Create lightweight capture points for rationales and assumptions at decision-heavy moments (merges, refactorings, hotfixes, deprecations).
    \item Support on-demand retrieval with provenance and freshness cues so maintainers can judge what is authoritative and what is outdated.
    \item Establish retention and curation policies that keep high-value knowledge while pruning redundant or obsolete context.
\end{itemize}
}

\textit{\textbf{Challenge V: Socio-Technical and Organizational Dependencies.}}  
Software development occurs within a socio-technical ecosystem that includes team practices, organizational policies, tooling, and communication structures. Dependencies often cross organizational boundaries: a system may rely on vendor-maintained APIs, open-source libraries, or internal services owned by different teams. Coordinating changes across these organizational interfaces can be as challenging as modifying the code itself. For instance, deprecating a shared API may require negotiating with multiple stakeholders, updating documentation, retraining support staff, and managing deployment pipelines. AI assistants have limited visibility into these organizational dynamics and cannot replace the negotiation, coordination, or alignment work central to socio-technical development.

\oppotunity{V}{
\begin{itemize}[label=-,leftmargin=10pt]
    \item Model socio-technical dependencies by linking technical artifacts to ownership, on-call responsibility, and cross-team interfaces.
    \item Integrate with planning and change-management workflows so coordination needs are detected early and reflected in work items.
    \item Provide coordination-aware recommendations that surface stakeholders, sequencing constraints, and communication checkpoints.
    \item Track coordination outcomes (breakages, delays, rework) to refine dependency maps and reduce repeated friction over time.
\end{itemize}
}


\subsection{Intelligent Software Development: AI-Assistant Perspective}
AI coding assistants can reduce routine effort by automating repetitive tasks, drafting code changes, supporting debugging, and assisting with compliance or performance checks. These benefits apply to both development and maintenance activities, where assistants can accelerate feature implementation, streamline refactoring, and reduce overhead for well-scoped work. However, effective use introduces its own challenges, including limitations in system-level context, reasoning across large codebases, and maintaining trustworthy behavior under human oversight. The following challenges illustrate opportunities and constraints when applying AI assistants in complex software development and maintenance.

\textit{\textbf{Challenge VI: Task Specification.}}  
Effectively leveraging AI assistants requires translating high-level development and maintenance objectives into structured, well-defined tasks. Poorly specified tasks can lead to incorrect or unsafe suggestions. For example, instructing an assistant to implement a new feature in a payment module requires specifying exact inputs and outputs, constraints such as transaction atomicity, error-handling requirements, and the relevant system context including dependent services or legacy APIs. Without careful task specification, the assistant may generate code that violates design constraints or business rules, introduces subtle bugs, or causes regressions. This challenge is compounded because many requests arrive informally, such as vague feature requests (``add support for international payments'') or issue tickets (``the system is slow'') that lack sufficient detail. Translating such requirements into actionable tasks often requires domain knowledge and contextual interpretation.

\oppotunity{VI}{
\begin{itemize}[label=-,leftmargin=10pt]
    \item Define structured task representations that specify intent, inputs/outputs, constraints, acceptance criteria, and required context.
    \item Provide interactive requirement elicitation that turns informal tickets into checkable task specifications with minimal extra burden.
    \item Attach validation hooks to task specifications (tests to run, invariants to preserve, interfaces to respect) before code is generated or changed.
    \item Close the loop by recording outcomes (accepted changes, rejected suggestions, failure modes) to improve future task formulation.
\end{itemize}
}

\textit{\textbf{Challenge VII: The Context Gap in AI Coding Tools.}}  
Many software tasks, whether development or maintenance, require awareness of system-level context, module interdependencies, and historical design decisions. Current AI coding tools (e.g., Cursor, GitHub Copilot) often rely on \textit{context engineering}, manually selecting and injecting relevant files into the prompt, to approximate this awareness. However, this approach faces two barriers in complex systems.
First, the \textbf{Fallacy of Infinite Context}: Simply increasing the context window does not guarantee useful reasoning. The ``Lost in the Middle'' phenomenon~\cite{liu2024lost} shows that performance degrades when relevant information is buried in noise, and raw repository data contains substantial noise (e.g., diffs and merge commits).
Second, the \textbf{Spatiotemporal Entanglement}: Software context is not just text; it is \textit{scattered} spatially (coupled across modules) and \textit{tangled} temporally (intent lives in past commits). Standard retrieval based on semantic similarity often misses these links, for example retrieving a function implementation but missing the configuration file that dictates its behavior, or the issue ticket that explains \textit{why} a particular logic exists.
Without a structured, version-aware context layer, assistants may produce plausible but architecturally invalid changes. \codedigitaltwin targets this gap by organizing artifacts and rationales into a persistent, curated knowledge layer.

\oppotunity{VII}{
\begin{itemize}[label=-,leftmargin=10pt]
    \item Expand context from text snippets to dependency-aware context, including call graphs, data flows, configuration bindings, and runtime interfaces.
    \item Implement context selection as a pipeline with explicit signals (relevance, authority, freshness, and coverage) rather than relying on ad hoc file picking.
    \item Apply systematic filtering and compression to remove noise (redundant diffs, generated artifacts, irrelevant history) while preserving key constraints.
    \item Support evolution-aware reasoning by surfacing what changed and why, including prior fixes, regressions, and compatibility constraints.
    \item Maintain a persistent, curated, version-aware context layer (for example, \codedigitaltwin) so retrieval remains stable as the codebase evolves.
\end{itemize}
}

\textit{\textbf{Challenge VIII: Non-Functional and Operational Constraints.}}  
Development often must satisfy non-functional requirements such as performance, security, reliability, and regulatory compliance. AI assistants may generate syntactically correct code, but ensuring that changes respect system-wide constraints requires validation. For example, a suggestion to optimize a database query might improve runtime efficiency but increase memory usage or bypass existing access controls. Similarly, modifying authentication logic could unintentionally weaken security by disabling multi-factor verification. Operational constraints further complicate the picture: code changes may need to comply with strict deployment schedules, high-availability requirements, or resource constraints in production environments (such as limited memory in embedded systems). Assistants typically lack direct visibility into these operational properties when they are not represented in accessible artifacts. Combining assistant suggestions with automated verification, monitoring, performance regression testing, and compliance checks is critical to safely address these constraints.

\oppotunity{VIII}{
\begin{itemize}[label=-,leftmargin=10pt]
    \item Represent non-functional requirements as explicit constraints (policies, budgets, invariants) that can be referenced and checked.
    \item Integrate pre-flight validation that runs tests, static analyses, and compliance checks aligned with those constraints.
    \item Surface trade-offs and risk signals (performance versus memory, security versus usability) with evidence rather than generic warnings.
    \item Feed operational telemetry and regression signals back into constraint definitions to keep them current.
\end{itemize}
}

\textit{\textbf{Challenge IX: Trust and Human Oversight.}}  
Even when AI assistants perform well, fully trusting them to autonomously modify production systems is risky. Minor mistakes can propagate rapidly, introducing regressions, security vulnerabilities, or compliance violations. For instance, an assistant might refactor code in a way that passes existing tests but breaks undocumented invariants, alters timing behavior critical for distributed coordination, or introduces performance bottlenecks under high load. In safety-critical domains such as healthcare or aviation, these risks are amplified because even small errors can lead to catastrophic outcomes. Moreover, assistant outputs are not always accompanied by evidence that is easy to audit, making it difficult for developers to assess the rationale behind suggestions or to build confidence in their correctness. Human oversight remains indispensable, not only for validation but also for incorporating tacit knowledge and system-level judgment that assistants cannot infer. Effective oversight mechanisms include staged deployment pipelines, human-in-the-loop verification, selective code review, and rollback mechanisms.

\oppotunity{IX}{
\begin{itemize}[label=-,leftmargin=10pt]
    \item Provide provenance for suggestions by linking them to dependencies, constraints, and supporting artifacts used in the reasoning.
    \item Communicate uncertainty and expected impact in a way that maps to developer actions (what to review, what to test, what could break).
    \item Integrate staged verification and rollout mechanisms (reviews, canaries, rollbacks) as a default part of assistant-supported changes.
    \item Preserve auditability by recording decisions, validations performed, and rationale for accepted or rejected suggestions.
\end{itemize}
}

\textit{\textbf{Challenge X: Intelligent Assistant Design.}}  
Developers require AI-assistant tools that can support development and maintenance activities, including automated debugging assistants, refactoring advisors, code summarizers, and documentation generators. Designing such assistants goes beyond surface-level code completion: they must combine code understanding, system reasoning, context modeling, and domain-specific knowledge. For instance, a tool recommending safe API migrations must account for historical usage patterns, module dependencies, backward compatibility requirements, and runtime constraints. Similarly, an intelligent debugging assistant might need to trace execution across microservices, identify concurrency issues, and provide explanations aligned with architectural documentation. Building these assistants also requires integration with developer workflows, such as version control systems, CI/CD pipelines, and issue trackers. Without such integration, assistant tools risk being isolated point solutions that add little value in real development contexts. Achieving robust, context-aware, and developer-centric assistant design remains a key challenge at the intersection of software engineering and AI.

\oppotunity{X}{
\begin{itemize}[label=-,leftmargin=10pt]
    \item Create modular assistants that combine code analysis, historical context, and domain knowledge for distinct development and maintenance tasks.
    \item Support iterative workflows that plan, act, and verify through tool integration (compilers, linters, analyzers, tests, CI, and issue-tracker queries).
    \item Maintain persistent, version-aware project memory (design rationales, constraints, past fixes) together with short-term session memory to reduce repeated context engineering.
    \item Enable cross-artifact reasoning that traces dependencies and proposes safe refactorings or migrations with evidence (tool outputs, affected files, and rollback plans).
    \item Integrate naturally with existing workflows, including version control, code review, CI/CD, and project tracking.
\end{itemize}
}

\textit{\textbf{Challenge XI: Realistic Evaluation and Benchmarking.}}  
Evaluating the effectiveness of AI assistants in development tasks requires benchmarks that reflect real-world complexity, including large legacy codebases, multi-module dependencies, evolving system requirements, and cross-team interactions. Current benchmarks often fall short of this need. For example, SWE-bench~\cite{jimenez2023swe}, developed by the AI community and widely adopted in the SE community, measures assistants' ability to resolve issues in open-source repositories. While useful for early exploration, it captures only a narrow slice of real development: most tasks are small-scale, focused on localized bug fixes, and lack the historical, architectural, and organizational complexity of large systems. It does not account for non-functional constraints, undocumented invariants, or undocumented knowledge loss, which are central challenges in practice. For instance, fixing a performance regression in a legacy banking platform may require balancing compliance rules, maintaining backward compatibility, and integrating with decades-old APIs, which are scenarios not captured in current benchmarks. Developing datasets and evaluation methodologies that incorporate such dimensions is therefore essential but remains a difficult and ongoing research task.

\oppotunity{XI}{
\begin{itemize}[label=-,leftmargin=10pt]
    \item Construct evaluation datasets that include design decisions, historical evolution, organizational interactions, and large-scale dependencies.
    \item Build benchmark tasks that require preserving non-functional and operational constraints, not only producing compiling patches.
    \item Use multi-dimensional metrics that capture correctness, safety, effort reduction, and downstream impact (tests, performance, regressions).
    \item Evaluate assistants longitudinally on evolving systems to measure robustness under version changes, shifting requirements, and accumulated context.
\end{itemize}
}

\section{\codedigitaltwin: A Living Knowledge Infrastructure}

The challenges and opportunities in complex software development highlight a recurring root cause: many critical engineering decisions depend on \textit{\textbf{tacit knowledge}} that is rarely explicit, frequently scattered, and continuously evolving. To work effectively in long-lived systems, AI assistants must reason beyond isolated code snippets, incorporating low-level artifacts (e.g., functions, files, modules), architecture and dependencies, change history, contextual decisions, and the design rationales and constraints behind existing solutions. This motivates an \textit{infrastructure layer} built on top of the codebase that captures and serves tacit knowledge as actionable context, separating long-term knowledge engineering from task-time context engineering.

\begin{figure}
\centering
\includegraphics[width=\linewidth]{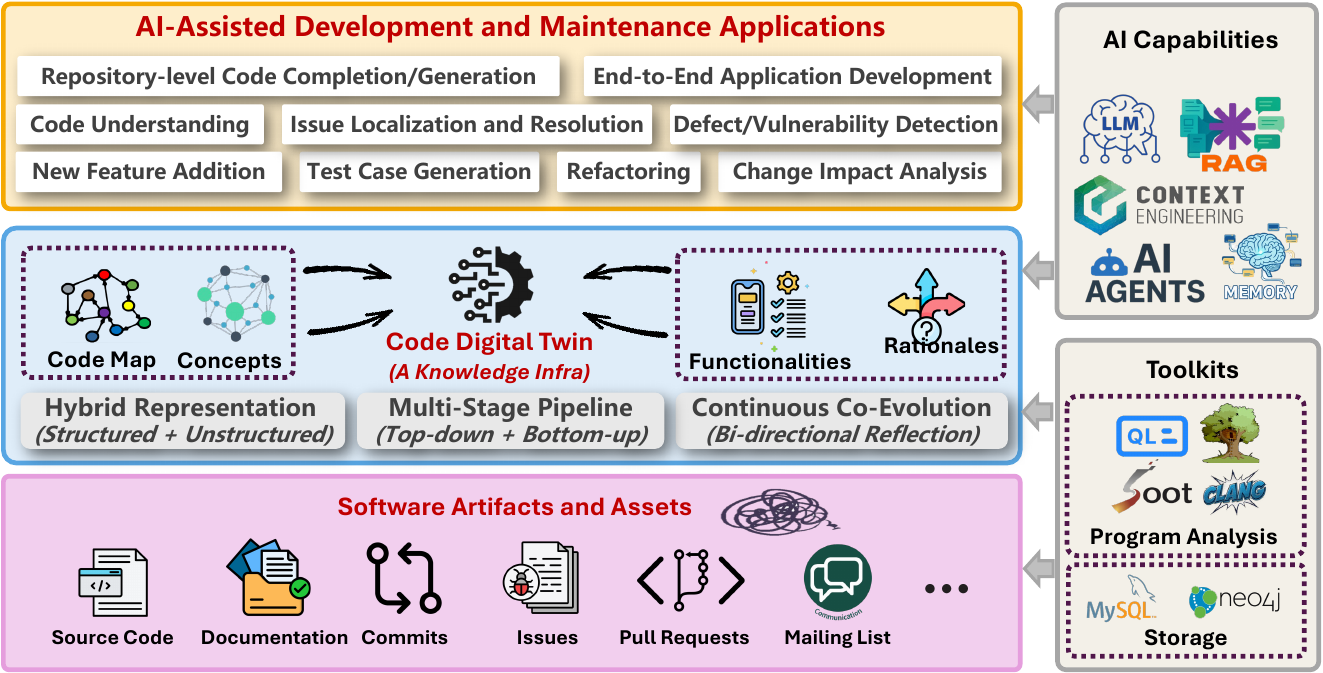}
\caption{Overview of the \codedigitaltwin landscape.}
\label{fig:overview}
\end{figure}

Our vision is \textit{\textbf{\codedigitaltwin}}, depicted in Figure~\ref{fig:overview}, a living knowledge infrastructure that couples a \textit{physical layer} of software artifacts with a \textit{conceptual layer} that encodes system intent and evolution. Inspired by the digital-twin paradigm in manufacturing, \codedigitaltwin provides a co-evolving model that makes tacit knowledge explicit, organizes it with hybrid representations (structured knowledge graphs and lightweight knowledge cards, plus supporting unstructured textual explanations), and supports collaborative curation by developers. As an infrastructure layer, \codedigitaltwin enables AI assistants and developers to assemble comprehensive, structured context for both software maintenance tasks (e.g., issue localization and change impact analysis) and iterative development tasks (e.g., integrating new features into an existing codebase).
In this paradigm, AI assistants act as context-aware collaborators guided by explicit system knowledge, while human judgment remains central.

\subsection{Definition}

The \codedigitaltwin infrastructure builds upon three foundational dimensions: \textit{physical software artifacts and related assets}, \textit{conceptual knowledge elements}, and an \textit{integrated twin model} that connects the two through explicit links and provenance. Together, these dimensions establish a version-aware, evolving knowledge layer that stays aligned with ongoing software evolution.

\subsubsection{``Physical'' Software Artifacts, Assets, and Their Traceability}
This dimension covers the tangible elements of a software system that developers and tools manipulate during development and maintenance, together with the records that describe how these elements evolve. At its core are \textit{\textbf{source artifacts}} such as functions/methods, classes, packages/modules, scripts, and configuration files. Surrounding them are \textit{\textbf{build and deployment artifacts}} (e.g., compiled binaries, container images, CI/CD pipeline definitions, infrastructure-as-code, and deployment scripts), \textit{\textbf{testing artifacts}} (unit/integration tests, test data, and coverage reports), and \textit{\textbf{third-party dependencies}} (libraries, frameworks, SDKs, and services). Complementary assets include \textit{\textbf{documentation and specifications}} (requirements, API references, architectural diagrams, READMEs), \textit{\textbf{issue and change records}} (bug reports, feature requests, release notes, and changelogs), and \textit{\textbf{discussion threads}} (e.g., pull-request reviews, design proposals, mailing lists, and chat logs).

To support longitudinal reasoning, \codedigitaltwin explicitly models \textit{\textbf{artifact traceability}}. Artifacts are anchored to their evolution and supporting records, including versions, commits, branches, pull requests, issues, and referenced documents. This enables queries such as which artifacts implemented a functionality at a given revision, which changes introduced a constraint, and which discussions justify a design decision. For online systems, the artifact layer may also incorporate \textit{\textbf{operational data}} such as logs, metrics, traces, and performance profiles, linked back to the corresponding deployed artifacts and configurations when available.

\subsubsection{``Conceptual'' Knowledge Elements and Their Relationships}
While artifacts embody concrete resources, software systems also rely on conceptual knowledge that provides intent, structure, and interpretability. Three categories of knowledge elements are particularly important.

\textit{\textbf{Domain concepts}} capture the fundamental abstractions that shape how a system is designed and reasoned about. They include concepts such as operating system primitives in kernel development, communication protocols in distributed systems, regulatory requirements in financial applications, and clinical standards in healthcare platforms. Without these abstractions, it is difficult to contextualize low-level code, understand why certain functionalities exist, or ensure that modifications remain aligned with business and compliance needs. For example, updating a distributed database without understanding consensus protocols risks introducing subtle consistency violations.

\textit{\textbf{Functionalities and responsibilities}} describe the core capabilities of a system and how responsibilities are allocated across components. These capabilities include workflows such as user authentication, transaction processing, or recommendation generation, as well as cross-cutting concerns like logging, error handling, caching, and synchronization. In complex systems, a single functionality is often implemented across multiple modules or services, with implicit boundaries, ownership, and interfaces. Capturing these mappings helps developers trace requirements back to code, reason about dependencies, and assess the broader implications of a change. For instance, a seemingly local modification to authentication logic may propagate across payment services, auditing workflows, and compliance subsystems.

\textit{\textbf{Rationales and constraints}} encode the decision-making logic underlying system design and implementation. They provide explanations for why a specific structure, algorithm, or interface was chosen, capturing trade-offs, constraints, and contextual reasoning. Rationales are often distributed across informal discussions, commit messages, or architectural notes rather than documented systematically. For example, a rationale might explain why a performance optimization was applied only in certain workflows due to consistency constraints, or why a legacy API was preserved to maintain backward compatibility. Recording these rationales is critical for avoiding repeated mistakes, understanding historical constraints, and guiding safe evolution.

Beyond the element types, \codedigitaltwin represents their \textit{\textbf{relationships}} explicitly so that reasoning and retrieval can follow structured links. In addition to artifact traceability, \codedigitaltwin makes (i) the \textit{concept--functionality--responsibility} \textit{\textbf{skeleton}} explicit and (ii) the \textit{\textbf{rationale and constraint links}} that explain and restrict this skeleton explicit. The skeleton captures how domain concepts are operationalized by concrete functionalities and how each functionality is realized through explicit responsibilities allocated across components. The rationale and constraint links preserve why the skeleton takes its current shape and what properties it must satisfy.

To make this systematic, we use a small, typed relationship vocabulary.
(1) \textit{operationalized-by} (concept $\rightarrow$ functionality) and \textit{requires} or \textit{uses} (functionality $\rightarrow$ concept) to capture the concept--functionality bridge.
(2) \textit{decomposes-to} and \textit{depends-on} (functionality $\rightarrow$ functionality) to capture functional structure and inter-functionality dependencies.
(3) \textit{has-responsibility} (functionality $\rightarrow$ responsibility) and \textit{assigned-to} (responsibility $\rightarrow$ component or boundary) to capture responsibility allocation.
(4) \textit{constrained-by} (concept, functionality, responsibility, or interface $\rightarrow$ constraint) to record invariants and non-negotiable requirements.
(5) \textit{justified-by} (allocation, interface, or design choice $\rightarrow$ rationale) to preserve trade-offs and decision context, with explicit pointers to supporting evidence.
These relations are version-aware through links to the corresponding artifacts and change records, enabling the twin to answer questions that require both intent and history.

Together, domain concepts, functionalities and responsibilities, and rationales and constraints form the conceptual scaffolding that complements physical artifacts and enables meaningful interpretation, extension, and development and maintenance of complex software systems.

\subsubsection{The \codedigitaltwin Model}
The \codedigitaltwin integrates artifacts and knowledge elements into a unified, continuously evolving model of a software system.

The model starts from a \textit{\textbf{code and artifact map}} that organizes the system's physical layer, including source artifacts, configurations, tests, build and deployment artifacts, and their dependencies. This map provides a structured substrate for navigation and analysis (e.g., contains, defines, imports, calls, depends-on, and configured-by), and is anchored by artifact traceability to versions, commits, pull requests, issues, and relevant documents.

Built on top of this substrate is an \textit{\textbf{functionality-oriented skeleton}} that focuses on the conceptual layer: \textit{domain concepts}, \textit{functionalities}, \textit{responsibility allocation}, and their \textit{interactions}. Rather than treating functionalities as isolated features, the skeleton captures how concepts are operationalized by workflows and cross-cutting concerns, how responsibilities are decomposed and assigned across components, and how functionalities depend on one another. By linking these conceptual structures back to the code and artifact map, \codedigitaltwin enables system-wide reasoning that connects intent to implementation. For example, it can expose a dependency subgraph that links an authentication functionality to its implementing modules, database schemas, configuration entries, and relevant tests, enabling maintainers to analyze the impact of a proposed change more holistically.

Complementing this skeleton is a layer of \textit{\textbf{rationale-centric explanations}} that ties artifacts and functionalities to their underlying decisions and constraints. For instance, if a caching mechanism was deliberately restricted to session-level data due to consistency considerations, the rationale layer preserves this justification alongside the corresponding code, documentation, and historical evidence (e.g., commits or discussions). This supports safer evolution by allowing developers and AI assistants to evaluate proposed optimizations in light of prior trade-offs and constraints.

Finally, the twin is characterized by \textit{\textbf{artifact-knowledge reflection and co-evolution}}, which ensures that knowledge remains synchronized with the evolving software system. As modules are refactored, new features are introduced, or dependencies are retired, the twin updates corresponding functionalities, rationales, constraints, and relationships, while preserving traceability to artifacts and histories (e.g., versions, commits, issues, and discussions). For example, introducing a new payment gateway would trigger updates to functionality records, revisions of responsibility allocation and dependency mappings, and links to the design decisions that motivated the change. This continuous alignment makes the twin a living, actionable representation of the system, rather than static documentation. It supports traceability, risk assessment, impact analysis, and collaborative reasoning between developers and AI assistants.

\subsection{Methodology and Roadmap}
To realize the \codedigitaltwin infrastructure, we present a systematic methodology alongside a research roadmap. The methodology operationalizes \codedigitaltwin as an always-available knowledge layer that is constructed from heterogeneous project artifacts, linked with traceability, and incrementally updated as the codebase evolves.

\subsubsection{Representation Schema}
To capture the breadth and depth of software artifacts and knowledge elements, \codedigitaltwin adopts a hybrid knowledge stack. \textit{\textbf{Structured representations}}, including knowledge graphs, frames, and cards, encode entities and relationships among artifacts, concepts, functionalities, responsibilities, and rationales. For example, a graph can express that a caching component depends on a database subsystem, that both implement a transaction-processing workflow, and that a specific constraint motivated a particular design choice. Complementing this, \textit{\textbf{unstructured representations}} preserve supporting evidence summarized and fused based on multi-source information such as commit messages, issue discussions, and design documents, which provide nuance and justification that cannot always be reduced to rigid schemas.

To make the representation actionable, we refine the schema along two complementary dimensions that align with the model definition: a \textit{code and artifact map} for the physical layer, and a \textit{functionality-oriented skeleton} plus \textit{rationale spine} for the conceptual layer.

\textit{\textbf{Artifact schema (Code and artifact map with traceability).}} The artifact layer is organized as a \textit{code and artifact map}, implemented as a typed \textit{code map} that covers files, functions/methods, classes, modules/packages, configuration entries, tests, build and deployment artifacts, and runtime components when available. Typical relations include structural edges (contains, defines, imports), behavioral edges (calls, reads-writes, depends-on), and environment edges (configured-by, built-into, deployed-as). In addition, \textit{artifact traceability} anchors these nodes and edges to histories and external records, including versions, commits, pull requests, issues, and design documents.

\textit{\textbf{Knowledge schema (Functionality skeleton and rationale spine).}} The knowledge layer captures (1) \textit{domain concepts}, (2) \textit{functionalities and responsibility allocation} across artifacts, and (3) \textit{rationales and constraints} that explain trade-offs and invariants. The \textit{functionality skeleton} is defined by domain concepts and functionality responsibilities, grounded in the code map through links such as implements, owns, and depends-on. Complementing it, a \textit{rationale spine} attaches explanations and constraints to the skeleton with evidence links.

Operationally, the representation layer can be designed around three concrete outputs: (1) a code and artifact map with artifact traceability links, (2) knowledge graphs and concise cards that summarize domain concepts, functionality responsibilities, and rationales with pointers to evidence, and (3) an evidence store of source text fragments anchored to artifact versions. This combination enables AI assistants to retrieve structured context while keeping explanations grounded in traceable evidence.

\textit{\textbf{Roadmap:}} Future research should focus on three directions. First, developing scalable mechanisms to integrate heterogeneous structured sources, such as domain-specific ontologies, dependency graphs, and architectural models, with unstructured corpora like documentation and mailing lists. Second, designing hybrid representation techniques that tightly link structured entities with extracted textual knowledge, ensuring both formal relationships and contextual explanations are accessible in a unified view. Third, addressing ambiguity and evolution by incorporating uncertainty-aware representations, knowledge versioning, and mechanisms for recording design trade-offs, thereby maintaining a coherent and up-to-date digital twin.

\subsubsection{Construction Pipeline}
The twin is assembled through a multi-stage knowledge mining pipeline that combines LLM-assisted extraction with static and dynamic program analysis. The pipeline is designed to produce version-aware, traceable knowledge, rather than one-off summaries.

\textit{Stage 1: Code and Artifact Map Construction.} This stage constructs the foundational representation of the software's physical layer. It builds a typed code map from static and dynamic analysis (e.g., module hierarchies, call graphs, dependency graphs, configuration references, and test structure) and then anchors code map entities to traceability sources, including versions, commits, pull requests, issues, and relevant documents. The result is a version-aware artifact substrate that later stages can reliably ground on.

\textit{Stage 2: Functionality-Oriented Skeleton Knowledge Extraction.} This stage builds the functionality skeleton, consisting of domain concepts and functionality responsibilities, and grounds it in the code map. Top-down, schema-guided extraction from documentation and specifications identifies domain concepts, user-facing features, constraints, and intended responsibility boundaries. Bottom-up, program analysis and LLM-assisted summarization and abstraction identify functionality implementations, cross-cutting concerns, and responsibility allocation across artifacts. The output is a structured skeleton that maps concepts and responsibilities to concrete artifacts and their dependencies.

\textit{Stage 3: Rationale-Centric Explanatory Knowledge Enrichment.} This stage enriches the skeleton with a rationale spine by extracting decisions, trade-offs, constraints, and invariants from unstructured sources such as commit histories, issue discussions, pull requests, and design threads. Each rationale is stored with explicit links to supporting evidence and is attached to the corresponding concepts, responsibilities, and artifacts that it explains.

\textit{Stage 4: Artifact-Knowledge Reflection Construction.} This stage constructs bidirectional links between knowledge elements and artifacts, including links to artifact versions and change records. These links enable traceability, impact analysis, and context-aware assistance. For example, when a developer proposes a change to a module, the twin can surface the affected functionalities, the components responsible for them, and the rationales and constraints that shaped their current design.

\textit{\textbf{Roadmap:}} Advancing the construction pipeline requires improving extraction accuracy through model adaptation, task-specific prompting templates, and cross-validation with program analysis. Scalability must be addressed by implementing distributed and incremental extraction mechanisms capable of handling large, complex, and frequently updated systems. Finally, adaptability should be ensured through real-time synchronization, allowing the twin to reflect evolving software artifacts and maintain an accurate, trustworthy representation.

\subsubsection{Co-Evolution and Incremental Update}
A defining feature of \codedigitaltwin is continuous alignment with the evolving software system. As artifacts are added, modified, or removed, updates should propagate across functionalities, rationales, constraints, and dependency mappings. In practice, an incremental update cycle can be driven by change events (e.g., commits, pull requests, releases, or CI signals): (1) detect changed artifacts and impacted dependency neighborhoods, (2) update or regenerate the affected knowledge cards and graph edges, (3) refresh traceability links to the new versions and change records, and (4) run validation checks, such as schema conformance, link integrity, and consistency between extracted dependencies and program analysis.

For example, introducing a new feature involves recording the new functionality, linking it to implementing modules and configurations, capturing the associated rationale from discussions or change requests, and updating dependency graphs. Refactoring similarly triggers updates to responsibility allocation and constraints, while preserving historical evidence.

\textit{\textbf{Roadmap:}} Future research should focus on enhancing co-evolution capabilities by automating continuous synchronization with evolving codebases, integrating developer feedback, and maintaining high-quality, consistent knowledge. Techniques such as CI/CD integration, automated change detection, and semantic validation can be employed to track the impact of code changes on dependent artifacts and knowledge elements, ensuring the twin remains responsive and aligned with real-world software evolution.

\subsection{Application Paradigm: Infrastructure-Enabled AI Assistance}
\codedigitaltwin operationalizes AI-assisted software engineering by connecting downstream assistants to a version-aware knowledge substrate. The key idea is to treat the \textit{code and artifact map} as the navigable physical substrate, and to expose the \textit{functionality-oriented skeleton} and \textit{rationale spine} as structured, queryable semantics over that substrate. This enables (i) retrieval and context assembly that preserve dependency structure and responsibility boundaries, and (ii) interaction patterns where assistants propose changes and knowledge updates while developers validate, correct, and curate the twin as part of routine workflows.

\subsubsection{Infrastructure-to-Assistant Bridging}
The \codedigitaltwin infrastructure enables downstream AI-assisted workflows by exposing \textit{queryable handles} over artifacts, semantics, and history, and by providing operators that assemble task-ready context grounded in traceable evidence. Concretely, an AI assistant can (i) resolve a user request to artifact and knowledge identifiers, (ii) retrieve a version-scoped, dependency-preserving twin subgraph, and (iii) materialize this subgraph into a bounded context package whose contents and ordering reflect responsibility boundaries, interfaces, and constraints.

\parabf{1. Knowledge-Aware Retrieval (Twin-RAG).}
Standard RAG retrieves text chunks based on semantic similarity, which can sever code relationships (e.g., retrieving a function body but missing its interface definition, callers, or configuration entry points). \codedigitaltwin upgrades this to \textit{Twin-RAG} by retrieving a \textit{twin subgraph} that is explicitly constructed from the code and artifact map and then enriched by the functionality skeleton and rationale spine. Operationally, retrieval can follow a graph-first query plan: (i) \textit{entity resolution} maps a user request (e.g., ``fix retry behavior for payment failures'') to candidate functionalities, components, and entry points; (ii) \textit{version selection} pins the query to a specific revision and its traceability neighborhood; (iii) \textit{subgraph expansion} pulls a bounded neighborhood using typed edges (calls, depends-on, configured-by, tests, implements, has-responsibility, constrained-by, justified-by); and (iv) \textit{subgraph ranking} prioritizes nodes that sit on responsibility boundaries, public interfaces, or constraint-bearing paths.

As a result, retrieval returns not only code fragments, but also the minimal supporting neighborhood needed for correct edits: the defining interfaces, direct and transitive callers, relevant configuration and feature flags, affected tests, and the constraint and rationale links that delimit safe changes.

\parabf{2. Context Engineering via Structured Context Packages.}
Interactive AI-assisted coding depends on selecting the right artifacts for an AI assistant's context window. However, naively dumping files introduces noise and can degrade reasoning (the ``Lost in the Middle'' phenomenon~\cite{liu2024lost}). \codedigitaltwin supports context engineering by converting a retrieved twin subgraph into a \textit{structured context package} with an explicit manifest and token budget.

Concretely, a context package can include (i) the target slice of code and artifacts (entry points, changed functions/classes, immediate dependencies, and relevant configuration), (ii) interface and boundary summaries extracted from the skeleton (what responsibilities the target component owns and what it must not assume), (iii) attached constraint and rationale links with evidence pointers, and (iv) validation hooks such as the affected tests, invariants, and expected failure modes. The package ordering can be driven by the twin structure: interfaces and constraints first, then implementation details, then peripheral neighbors and evidence. This turns context selection into a controllable compilation step rather than ad hoc prompt assembly.

\parabf{3. Long-Term Memory for AI Assistants.}
Autonomous assistants can suffer from amnesia and treat every task as a new project. \codedigitaltwin provides a persistent \textit{Long-Term Memory} by maintaining the code and artifact map, functionality skeleton, and rationale spine as version-aware stores with traceability links. This supports longitudinal queries that combine intent and history, such as ``How did retry logic evolve across releases, which constraints were introduced, and which changes were rolled back?'' or ``Which design decision limits caching scope, and where is it enforced?''

In addition, the twin enables \textit{memory writeback} as part of normal workflows: when a change is accepted, the update cycle can automatically attach new evidence (commits, reviews, incidents) and refresh the affected knowledge cards and links. This makes the assistant's behavior stateful across sessions because future retrieval and context packaging reuse the updated skeleton and rationale spine rather than relying on ephemeral conversation history.

\textit{\textbf{Roadmap:}} Future research should focus on making the twin directly consumable by AI coding backends and IDE assistants through well-defined interfaces and operators. Key directions include (i) designing a twin query language that supports entity resolution, revision scoping, and bounded subgraph retrieval with typed edges and provenance, so that questions such as ``What breaks if I change X?'' return responsibility-aware impact neighborhoods rather than unstructured text; (ii) building context-package compilers that transform retrieved subgraphs into token-bounded, ordered manifests with explicit constraints, evidence pointers, and validation hooks; and (iii) defining safe writeback protocols that govern how assistants propose knowledge updates (new links, rationales, constraints) and how these updates are verified, versioned, and reviewed. Together, these advances move the twin from a passive store to an active substrate that supports reliable retrieval, context assembly, and stateful evolution during AI-assisted development.

\subsubsection{Human in the Loop}
The \codedigitaltwin evolves through a human-in-the-loop curation process that complements automated extraction and assistant-driven writeback. Developer and stakeholder input validates extracted knowledge, corrects boundary cases, and supplies context that automation may miss. This feedback loop can be integrated into routine workflows: developers can confirm or edit concept and functionality cards during reviews, attach missing evidence for rationales, and approve or reject knowledge update requests that modify the code and artifact map, the functionality skeleton, or the rationale spine.

Beyond explicit edits, the infrastructure can also leverage \textit{implicit feedback signals} produced during \textit{AI--human collaboration}. Examples include developers accepting or rejecting an assistant-proposed patch, selecting among alternative suggestions, correcting retrieved context, editing an auto-generated summary, or overriding a suggested responsibility boundary. These interactions can be logged as typed, structured events with provenance (e.g., pull request identifiers, review comments, and resulting commits) and compiled into controlled, reviewable knowledge updates. Concretely, such signals can drive (i) confidence calibration for entities, links, and extracted claims, (ii) refinement of boundaries and dependency edges when reviewers clarify ownership or interfaces, and (iii) creation or tightening of constraints when reviewers cite invariants, performance budgets, or backward-compatibility requirements. When signals reveal disagreement (e.g., conflicting rationales or incompatible responsibility allocations), the twin can surface them as curation tasks with linked evidence, rather than silently overwriting prior knowledge.

\textit{\textbf{Roadmap:}} Future research should enhance the synergy between human input, automated extraction, and assistant writeback. Key directions include designing feedback-aware interfaces that surface high-impact uncertainties, learning from implicit supervision signals without disrupting developer workflows, developing validation pipelines that combine evidence checks with consistency constraints, and implementing audit-friendly mechanisms for conflict resolution and knowledge versioning. These efforts ensure a resilient, collaboratively maintained digital twin that evolves in lockstep with the software system, supporting informed, context-aware decision-making for both developers and AI assistants.

\subsection{Preliminary Results}
To demonstrate the practical utility of \codedigitaltwin, we present two representative case studies: issue localization and application generation. In both cases, we focus on \textit{code artifacts only} and instantiate a code-centric twin model where extracted concepts and functionalities are linked, via explicit traceability, to concrete code elements (e.g., files and functions). These tasks illustrate the challenges of scaling AI assistants, powered by large foundation models, to complex, real-world software systems and show how functionality-oriented knowledge becomes more actionable when it is grounded in traceable code links.

\subsubsection{Issue Localization}
Issue localization, which involves identifying the precise code elements responsible for a reported defect, is a critical bottleneck in software maintenance. Large-scale repositories complicate this task due to concern mixing, where critical logic is buried within multi-purpose functions, and concern scattering, where related logic is dispersed across multiple files and modules. Existing AI-assistant approaches built on foundation models, such as mini-SWE-agent~\cite{yang2024swe}, often mislocalize issues due to unguided exploration and the absence of high-level conceptual context.

To address these limitations, we developed a knowledge extraction tool that leverages concept–functionality knowledge to abstract fine-grained code behaviors into high-level concerns. A concern represents a cohesive set of functionalities associated with a domain concept or feature, potentially spanning multiple files or modules. During localization, the most relevant concerns guide AI-assistant exploration, providing a structured, conceptual perspective aligned with human reasoning.
In this case study, each extracted concern is linked, through explicit traceability, to its implementing code elements (e.g., files and functions), so that assistants can retrieve and navigate the corresponding code neighborhoods rather than relying on free-form descriptions.

Experiments on the SWE-Lancer benchmark~\cite{lancer}, which contains 216 localization tasks drawn from a real-world repository exceeding 220 million lines of code, demonstrate substantial improvements. Using GPT-4o as the base model, the extracted knowledge increases Hit@k by over 22 percent and Recall@k by 46 percent across file- and function-level localization. Generalization tests across multiple models (GPT-4o, GPT-4o-mini, GPT-4.1) confirm consistent gains, with relative improvements ranging from 2.76 percent to 504.35 percent for Hit@1 and 2.83 percent to 376.13 percent for Recall@10. Ablation studies show that both conceptual term explanations and concern clustering contribute critically to performance, and manual annotations confirm the correctness, completeness, and conciseness of the extracted concerns. These results indicate that linking concept–functionality knowledge to traceable code elements enables AI assistants to reason more holistically, preserve architectural context, and navigate complex, scattered code more effectively.

\subsubsection{Application Generation}
End-to-end software generation with agentic AI assistants offers opportunities for automating complex development tasks, but prior approaches typically rely on linear, waterfall-style workflows or single-agent pipelines. This limitation reduces their effectiveness in large-scale, real-world software projects. While prior work has focused on small Python applications or simple web pages, more complex domains such as Android development involve intricate lifecycles, asynchronous tasks, diverse dependencies, and platform-specific APIs, which present additional challenges.

We propose an iterative, agentic approach inspired by feature-driven development that captures functionality-oriented knowledge at multiple levels. User requirements are decomposed into a set of user-valued features and structured into a Feature Map, a directed acyclic graph that explicitly models dependencies among features. Each node in the graph stores detailed knowledge about the feature, including business logic, design considerations, and code implementations. As development proceeds iteratively, this structured knowledge propagates through dependent features, ensuring context-aware guidance for AI assistants and maintaining coherence across the codebase. Note that, although the scenario involves end-to-end application generation, the iterative development approach inherently requires maintenance-related capabilities, such as integrating new features into existing codebases.
In this case study, the feature-level knowledge is grounded in code by linking each feature node to the relevant implementation units (e.g., classes, methods, and files), enabling traceable context reuse across iterations.

Evaluations on complex Android development tasks demonstrate the effectiveness of this approach. Compared with state-of-the-art agent frameworks built on foundation models, our feature-map-guided approach outperforms all baselines, including Claude Code, achieving a 56.8 percent improvement and relative gains of 16.0 to 76.6 percent across multiple base models. The structured propagation of concept–functionality knowledge enables AI assistants to understand inter-feature dependencies, maintain architectural and functional consistency, and handle large-scale iterative development, illustrating that knowledge-aware infrastructures are essential for scaling AI-assisted software generation beyond simple applications.

\subsubsection{Implications}
These case studies provide preliminary evidence that functionality-oriented knowledge in \codedigitaltwin becomes practically useful when it is \textit{linked back to code}. By capturing high-level concepts and functionalities and attaching explicit traceability links to concrete code elements, AI assistants can perform more accurate issue localization, manage complex dependencies in iterative development, and make more informed, context-aware decisions. Importantly, these results reflect a \textit{code-centric instantiation} of the broader \codedigitaltwin definition. In the future, we will extend these studies toward richer twin models that incorporate additional artifacts and longitudinal traceability, and evaluate \codedigitaltwin in more demanding scenarios that require extensive design considerations within large-scale software such as the Linux kernel.
\section{Related Work}
\parabf{AI-assisted development tools and context management.}
Tools such as GitHub Copilot, Cursor, and Claude Code illustrate a shift toward AI assistants that operate directly inside developer workflows~\cite{copilot, cursor, claudecode}. To support repository-scale tasks, recent work studies how to select, order, and compress prompt context~\cite{liu2024lost, shi2023large, xu2024retrieval}, and how to expose repository structure via indexes and graphs~\cite{liu2024codexgraph, ouyang2024repograph}. Most of these approaches treat context as \textit{static} and \textit{explicit}, by primarily reading what currently exists in the working tree. In contrast, \codedigitaltwin positions context as \textit{constructed} and \textit{maintained} by a living knowledge infrastructure that captures long-lived software knowledge and its evolution, so that task-time context can be compiled from curated, version-aware representations.

\parabf{Repository-level code generation with AI assistants.}  
Repository-level code completion involves generating code based on cross-file context within a repository. Two main techniques assist AI assistants in this task: Retrieval-Augmented Generation (RAG)~\cite{wu2024repoformer, liu2024stall+, zhang2023repocoder, liang2024repofuse}, which retrieves similar code snippets, and static analysis~\cite{agrawal2023guiding, wang2024teaching, liu2024stall+, liang2024repofuse, pei2023better}, which uses imported files or API lists. However, both approaches face challenges: retrieved snippets may not always help, and static analysis can be inaccurate, especially for dynamically typed languages. Despite advances in AI assistants, accurate repository-level code completion remains difficult due to cross-file context complexities and limitations of current techniques.


\parabf{Agentic software engineering for repository-level issue resolution.} 
Repository-level issue resolution is often framed as a search-and-edit loop over a large and evolving codebase. Recent studies propose agent-based methods~\cite{yang2024swe, zhang2024autocoderover, chen2024coder, liu2024marscode} and pattern-based methods~\cite{xia2024agentless}. Agent-based approaches (e.g., SWE-Agent~\cite{yang2024swe}) treat AI assistants as tool-using agents that navigate repositories, apply edits, and iteratively validate changes; pattern-based approaches (e.g., AgentLess~\cite{xia2024agentless}) separate retrieval/localization from editing. Benchmarks such as SWE-bench~\cite{jimenez2023swe} and follow-up efforts on longer-horizon tasks~\cite{lancer, deng2025swebenchproaiagents} highlight that success depends heavily on selecting the right context, maintaining coherent task state, and handling cross-file dependencies. These approaches typically build task context on demand and often lack a persistent, project-specific representation of higher-level responsibilities and their traceability to code. \codedigitaltwin complements them by providing an infrastructure for maintaining such long-lived knowledge and compiling task-time context packages grounded in traceable project representations.

\parabf{Repository comprehension via structured representations.}  
Repository comprehension combines code understanding with representations that make cross-file relations explicit. For example, MutableAI~\cite{MutableAI} converts codebases into Wikipedia-style articles with citations, and CODEXGRAPH~\cite{liu2024codexgraph} connects AI assistants to graph database interfaces derived from repositories so that they can query code structure. RepoGraph~\cite{ouyang2024repograph} and related studies~\cite{ma2024understand} further explore repository-level graphs for retrieval and comprehension. These systems demonstrate the value of wikis and graphs as intermediates, but they often focus on summarizing observed artifacts. \codedigitaltwin extends this direction by emphasizing a living knowledge infrastructure that connects repository structure to functionality-oriented knowledge (e.g., responsibilities and interactions) and supports controlled updates as the project evolves.

\parabf{Knowledge graphs for software engineering.}  
Researchers have developed various knowledge graphs to aid software development tasks. API KGs~\cite{icsme2018apicaveat,fse19apisummary,ase20apicomp,ase20apimisuse,icsme18docgen,icsme20docgen} summarize API knowledge, supporting tasks like migration and misuse detection. Software development concept KGs~\cite{saner17domainkg,fse2019glossary,wang2023xcos,TSE23CONCEPTLINK,jos2021automatic,icsme23kgxqr} create domain glossaries. Programming task KGs~\cite{fse22taskkg,icsme19taskkg} extract task attributes and relationships from tutorials, helping developers solve programming questions. Bug KGs~\cite{saner19codekg,su2021reducing,fse23crashkg} capture knowledge on bug causes and solutions. However, no knowledge base specifically summarizes design-related knowledge for long-term software maintenance tasks.
\section{Conclusion}
In this paper, we introduce \codedigitaltwin as a living knowledge infrastructure that bridges the gap between AI assistants and the tacit knowledge required for complex software development. \codedigitaltwin couples a physical substrate (a code and artifact map with traceability) with a conceptual substrate that makes the concept--functionality--responsibility skeleton explicit and records rationale and constraint links with evidence. We present a systematic methodology that integrates program analysis, LLM-assisted extraction, and human expertise to construct and continuously evolve this representation alongside ongoing software changes, enabling task-time context to be compiled from curated, version-aware knowledge rather than ad hoc prompts. Our preliminary results instantiate this vision in a code-only setting, where conceptual elements are explicitly linked back to files, functions, classes, and related implementation artifacts to support repository-scale reasoning.

Looking ahead, key challenges include scaling twin construction and updates to large, fast-evolving projects, governing knowledge quality and provenance, expanding from code-only instantiations to broader artifact and history traceability, and establishing evaluation protocols that measure how such infrastructures improve the reliability and maintainability of AI-assisted development.

\bibliographystyle{ACM-Reference-Format}
\balance
\bibliography{ref.bib}

\end{document}